\documentclass[a4paper]{jpconf}
\usepackage{graphicx}

\usepackage{citesort}

\begin{document}

\title{Constraints on the properties of the turbulent magnetic field around Geminga using HAWC measurements}

\author{Gwenael~Giacinti$^1$ and Rub\'en~L\'opez-Coto$^2$}

\address{$^1$ Max-Planck-Institut f\"ur Kernphysik, Postfach 103980, D-69029 Heidelberg, Germany}

\address{$^2$ Universit\`{a} di Padova and INFN, I-35131, Padova, Italy}

\ead{Gwenael.Giacinti@mpi-hd.mpg.de}

\begin{abstract}
We place constraints on the properties of the interstellar turbulence that surrounds Geminga pulsar, using the recent measurements from the HAWC Observatory in this region~\cite{Geminga_hawc}. We propagate very-high-energy electrons in realizations of 3D isotropic Kolmogorov or Kraichnan turbulence, calculate their $\gamma$-ray emission, and compare with HAWC measurements. We show that the measurements can be well fitted for both models of the turbulence and for reasonable values of its strength, $B_{\rm rms}$, and coherence length, $L_{\rm c}$. Our best fits are obtained for $B_{\rm rms} \simeq 3\,\mu$G and $L_{\rm c} \simeq 1$\,pc. Furthermore, the absence of strong asymmetries in the observed emission favours $L_{\rm c} \leq 5$\,pc.
\end{abstract}

%%%%%%%%%%%%%%%%%%%%%%%%%%%%%%%%%%%%%%%%%%%%%%%%%%%%%%%%%%%%%%%%%%%%%%%%%%%%%%%%%%
\section{Introduction}
%%%%%%%%%%%%%%%%%%%%%%%%%%%%%%%%%%%%%%%%%%%%%%%%%%%%%%%%%%%%%%%%%%%%%%%%%%%%%%%%%%

Extended gamma-ray emissions around Galactic sources of cosmic-rays (CR) can provide insights into the properties of the turbulent magnetic fields that surround these sources, see e.g.~\cite{2012PhRvL.108z1101G,2013PhRvD..88b3010G}. Such an emission has been recently detected by HAWC around Geminga, and is thought to be due to $\sim 100$\,TeV electrons and positrons diffusing and cooling around the pulsar~\cite{Geminga_hawc}. HAWC Collaboration measured the value of the diffusion coefficient of these electrons: $D_{\rm 100\,TeV} = (4.5 \pm 1.2 ) \times 10^{27}$\,cm$^2$\,s$^{-1}$ at 100\,TeV, and noted that its extrapolation to lower energies is about two orders of magnitude smaller than the Galactic average for CRs as deduced from B/C measurements. In the present paper, we put constraints on the turbulence within $\approx 25$\,pc from Geminga, using the measurements published in Ref.~\cite{Geminga_hawc}. We find that isotropic Kolmogorov or Kraichnan turbulence with a coherence length $L_{\rm c} \simeq 1$\,pc provide a good fit to the measurements~\cite{2018MNRAS.479.4526L}.

%%%%%%%%%%%%%%%%%%%%%%%%%%%%%%%%%%%%%%%%%%%%%%%%%%%%%%%%%%%%%%%%%%%%%%%%%%%%%%%%%%
\section{Method}
%%%%%%%%%%%%%%%%%%%%%%%%%%%%%%%%%%%%%%%%%%%%%%%%%%%%%%%%%%%%%%%%%%%%%%%%%%%%%%%%%%

We describe here how we produce the synthetic $\gamma$-ray maps that are compared with HAWC measurements in the next section. Instead of using the diffusion-loss equation, we propagate individual very-high-energy electrons (5000 for each map) in realizations of 3D isotropic Kolmogorov or Kraichan turbulence. This technique allows us to take into account effects that cannot be described properly within the standard isotropic diffusion approximation, such as highly anisotropic propagation of CRs along magnetic field lines. In particular, when most electrons escaping from a source are still located at distances smaller than $\approx L_{\rm c}$ from it, their $\gamma$-ray emission should highlight field lines around the source and, therefore, appear filamentary.

Measurements of the $\gamma$-ray spectrum of Geminga by Ref.~\cite{Geminga_hawc} shows that it follows a power law $dN/dE \propto E^{-2.34}$ between 8\,TeV and 40\,TeV. This is compatible with electrons being injected with a spectrum $dN_{\rm e}/dE \propto E^{-2.24}$ at the pulsar. In the simulations, we inject the electrons at the location of the source, which we assume to be point-like. The initial energies of these electrons are chosen between 40\,TeV and 500\,TeV and follow a power-law spectrum $\propto E^{-2.24}$. We take into account their synchrotron and inverse Compton energy losses during propagation, which are the dominant losses at the energies we consider, see Ref.~\cite{Diffusion_electrons_paper}. We neglect the effect of infrared and optical photons. Following Ref.~\cite{Moderski05}, the energy loss per time unit of an electron with energy $E_{\rm e}$ and placed in a magnetic field $B$ reads:
\begin{equation}
\left| \frac{dE_{\rm e}}{dt} \right| \simeq 2.53 \times 10^{-15} \, {\rm TeV/s} \, \bigg[ \left( \frac{B}{1\,\mu{\rm G}} \right)^{2} + \frac{10.1}{\left( 1 + E_{\rm e}/(99\,{\rm TeV}) \right)^{1.5}} \, \bigg] \left(\frac{E_{\rm e}}{{\rm 1\,TeV}}\right)^{2}\;.
\end{equation}

We generate the 3D turbulent magnetic fields in which we propagate these particles with the method presented in Ref.~\cite{Giacinti2012}. We test both isotropic Kolmogorov, $P(k) \propto k^{-5/3}$, and Kraichnan turbulence, $P(k) \propto k^{-3/2}$, with root-mean-square strengths in the range $B_{\rm rms} = (2 - 5)\,\mu$G and coherence lengths in the range $L_{\rm c} = (0.1 - 40)$\,pc. Trajectories are stopped once the particles reach 39\,TeV, because lower energy electrons do not contribute to the range of photon energies considered here. Our turbulence contains fluctuations down to scales smaller than the smallest electron gyroradius present in our simulations. Since HAWC measurements do not show any strong asymmetry in the emission, we do not add any large-scale magnetic field to our turbulence, see the discussion in Section~4. Since the typical cooling time of $\sim 100$\,TeV electrons is only $\sim 10$\,kyr, we assume that the electrons have been injected steadily on these time scales. Instead of injecting the electrons continuously in the simulation, we inject them at $t=0$, and record their momenta and positions at equally-spaced intervals in time, every $\Delta t$, and consider each recording as a new particle for the total emission. We verified that $\Delta t = 20$\,yr gives correct results. The gamma-ray emission is dominated here by the upscattering of CMB photons. We calculate it, using the \texttt{gamera}~\cite{GAMERA} and \texttt{edge}~\cite{Diffusion_electrons_paper} libraries. Finally, since the number of electrons in our simulations is much smaller than that present around Geminga in reality, we normalize our total $\gamma$-ray emission to that measured by HAWC.

%%%%%%%%%%%%%%%%%%%%%%%%%%%%%%%%%%%%%%%%%%%%%%%%%%%%%%%%%%%%%%%%%%%%%%%%%%%%%%%%%%
\section{Results}
%%%%%%%%%%%%%%%%%%%%%%%%%%%%%%%%%%%%%%%%%%%%%%%%%%%%%%%%%%%%%%%%%%%%%%%%%%%%%%%%%%

%%%%%%%%%%%%%%%%%%%%%%%%%%%%%%%%%%%%%%%%%%%%%%%%%%%%%%%%%%%%%%%%%%%%%%%%%%%%%%%%%%
\begin{figure*}
\begin{center}
\includegraphics[width=0.495\textwidth]{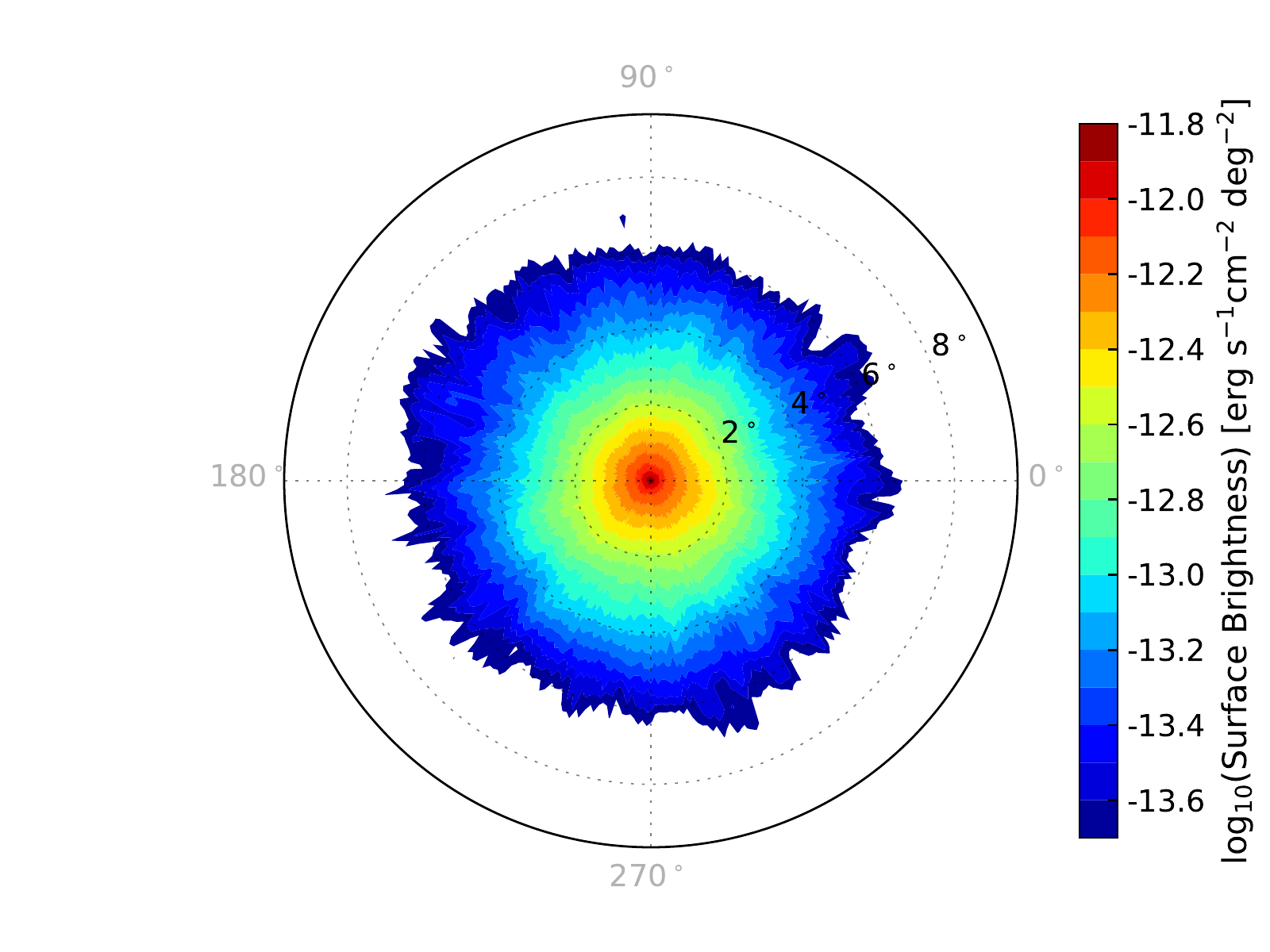}
\includegraphics[width=0.495\textwidth]{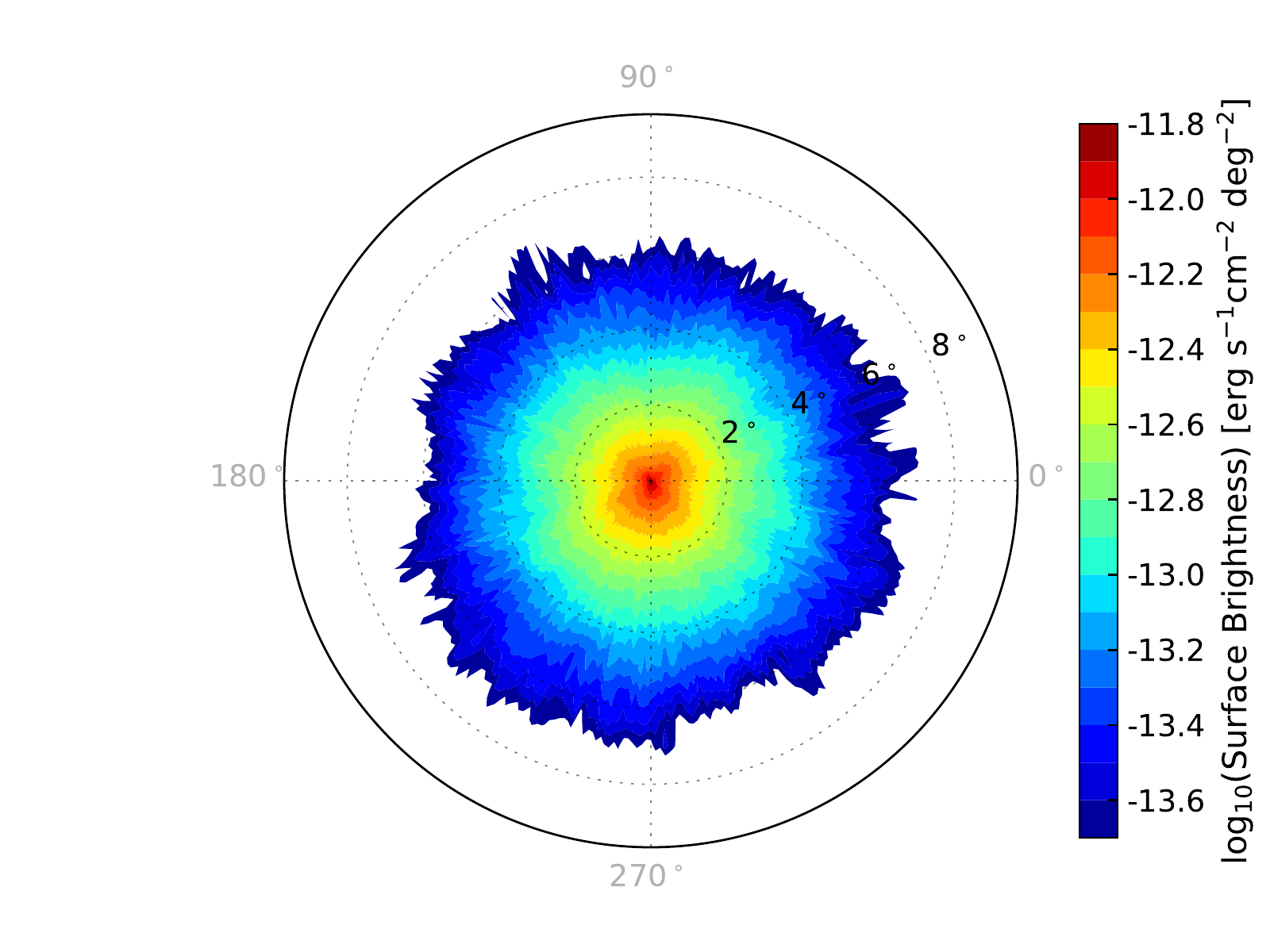}
\includegraphics[width=0.495\textwidth]{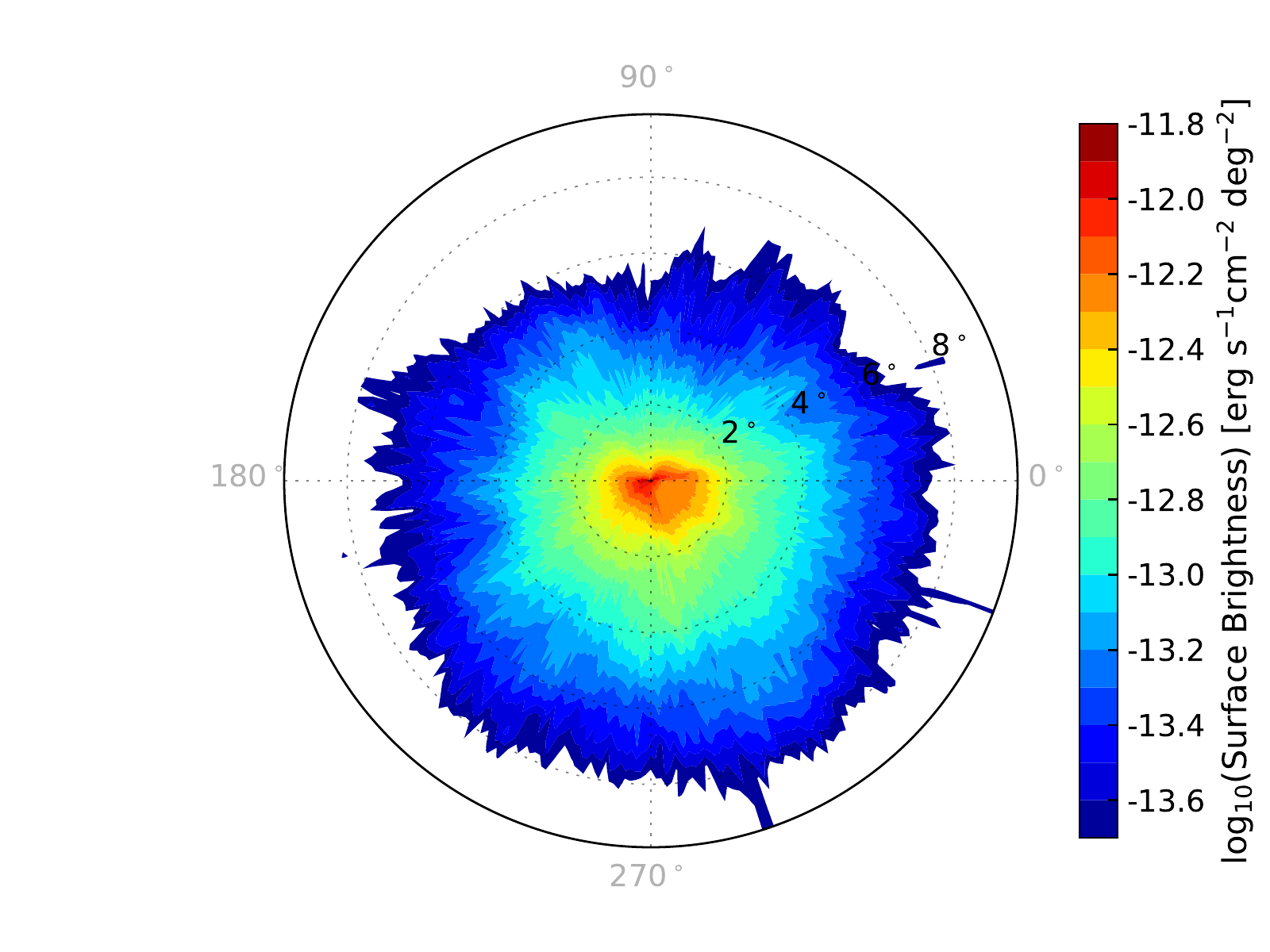}
\includegraphics[width=0.495\textwidth]{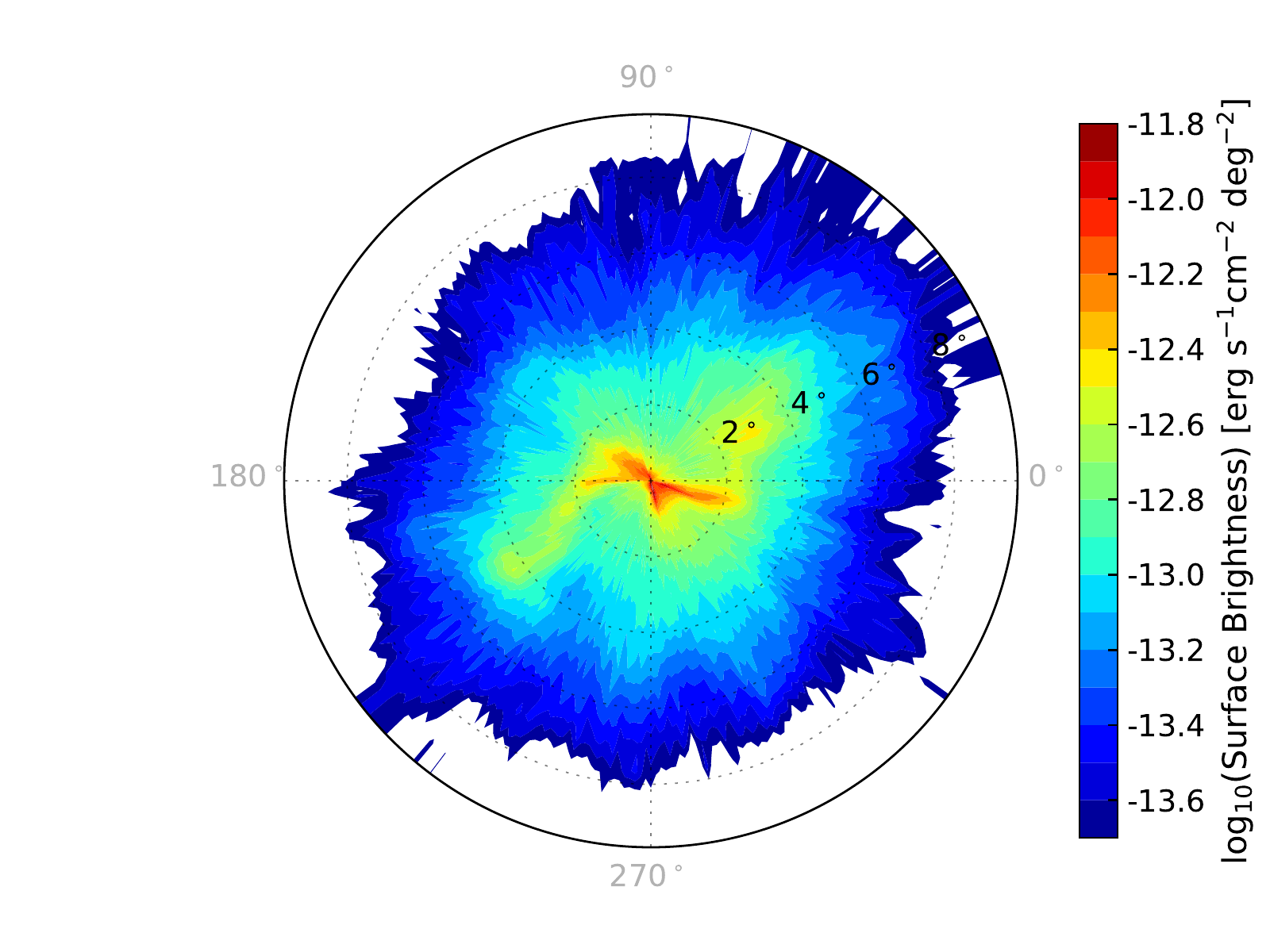}
\includegraphics[width=0.495\textwidth]{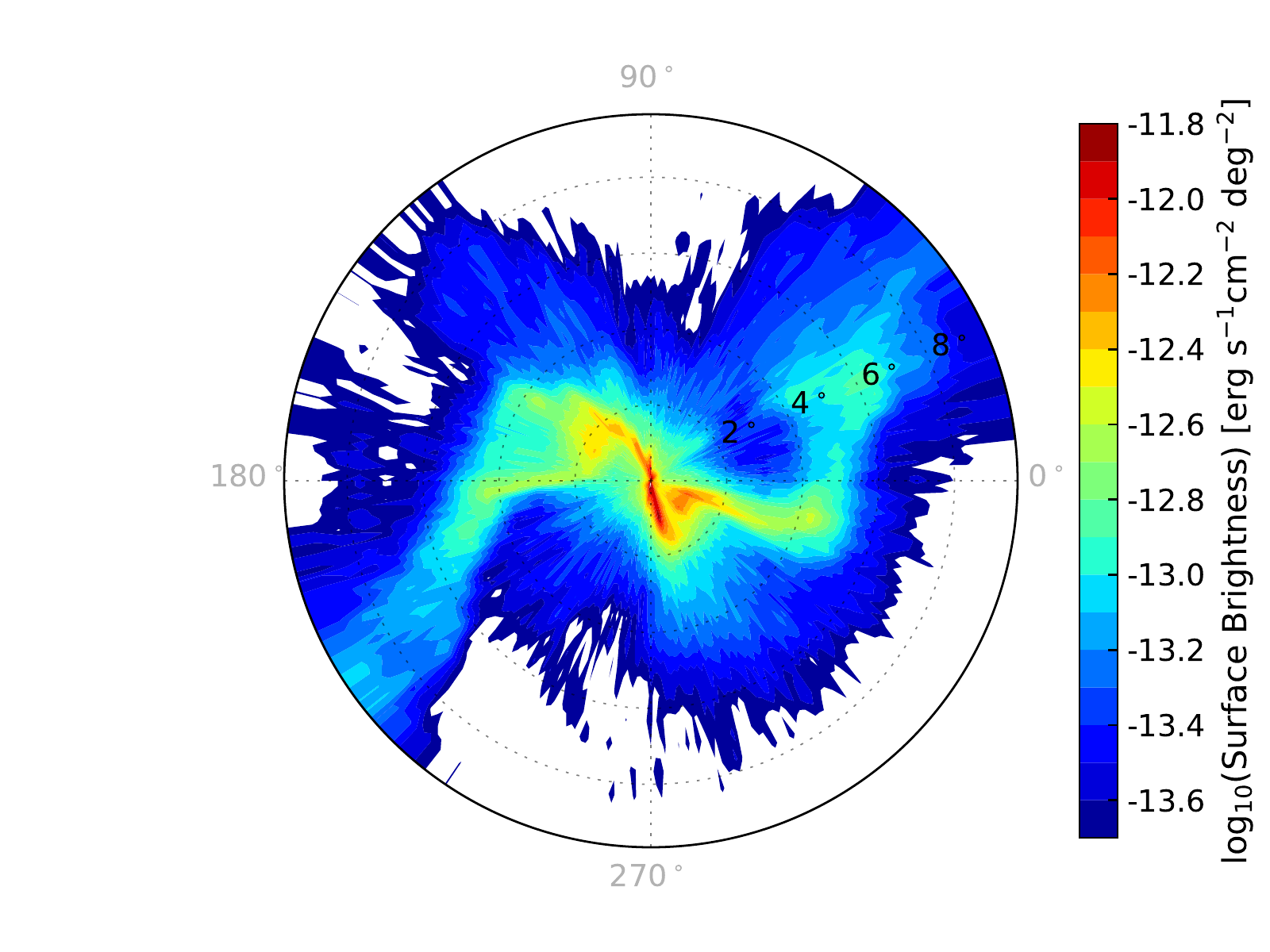}
\includegraphics[width=0.495\textwidth]{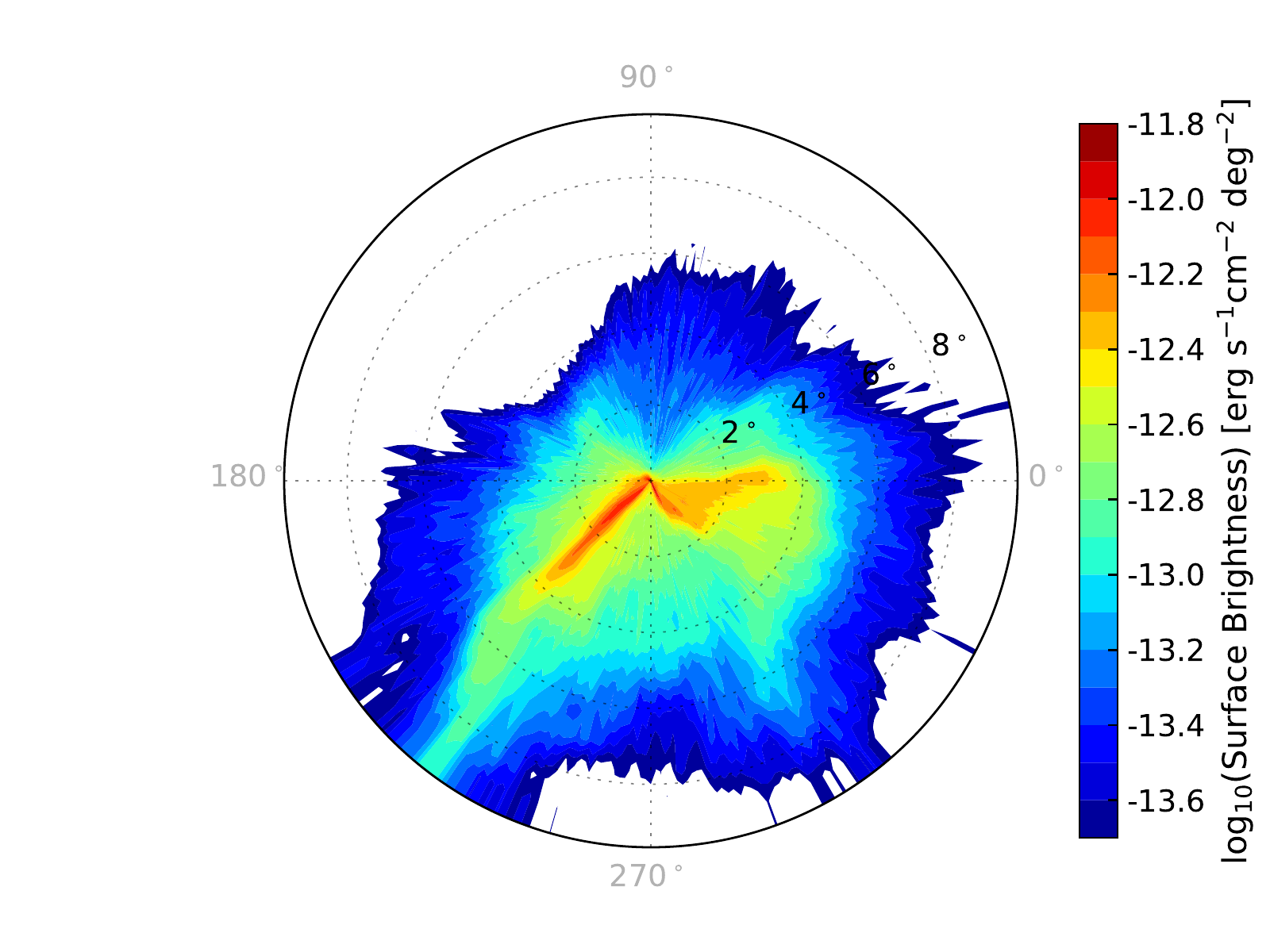}
\caption{\label{Fig1} Plots of the simulated gamma-ray surface brightness around Geminga (located at the centre of each plot), as viewed from the Earth. Electrons are injected continuously at the position of Geminga in given realizations of 3D isotropic Kolmogorov turbulence, with $B_{\rm rms} = 3\,\mu$G and $L_{\rm c}=0.25$\,pc (top left), 1\,pc (top right), 5\,pc (middle left), 10\,pc (middle right), 20\,pc (bottom left), or 40\,pc (bottom right). The polar angle is written in grey, and the angular distance from the pulsar in black.}
\end{center}
\end{figure*}
%%%%%%%%%%%%%%%%%%%%%%%%%%%%%%%%%%%%%%%%%%%%%%%%%%%%%%%%%%%%%%%%%%%%%%%%%%%%%%%%%%

In Fig.~\ref{Fig1}, we plot the simulated $\gamma$-ray surface brightness, as viewed from the Earth, in a $10^{\circ}$-radius region around Geminga. The pulsar is located in the centre of each plot. In all six panels, we use isotropic Kolmogorov turbulence with root-mean-square strength $B_{\rm rms} = 3\,\mu$G. In each panel, we set the coherence length to a different value: $L_{\rm c}=0.25$\,pc (top left panel), $L_{\rm c}=1$\,pc (top right), $L_{\rm c}=5$\,pc (middle left), $L_{\rm c}=10$\,pc (middle right), $L_{\rm c}=20$\,pc (bottom left), and $L_{\rm c}=40$\,pc (bottom right). Red and yellow regions are the brightest ones, see the colourbars on the side of the panels for the values of the surface brightness. The emission clearly becomes increasingly asymmetric (with respect to rotations around the position of the pulsar) for increasing values of $L_{\rm c}$. This is due to the fact that neighbouring magnetic field lines close to the source remain close to one another typically up to distances $\approx L_{\rm c}$ from the source. For small values of $L_{\rm c}$ (approximately $\leq 5$\,pc in these plots), magnetic field lines are tangled on scales that are significantly smaller than the size of the $\gamma$-ray emitting region. Therefore, even if CRs follow magnetic field lines individually, the resulting $\gamma$-ray emission around the source looks quite symmetric to the observer. In contrast, for larger values of $L_{\rm c}$ (approximately $\geq 10$\,pc in these plots), $L_{\rm c}$ is not small compared with the size of the $\gamma$-ray emitting region, and the bulk of escaping CRs is mostly confined in a magnetic flux tube of length $\leq L_{\rm c}$ around the source. The filamentary structures visible in $\gamma$-rays highlight the local magnetic field lines. Therefore, the presence or absence of asymmetries in the emission can be used to put constraints on the coherence length of the turbulence. Since no strong asymmetry has been detected by HAWC yet, the results in Fig.~\ref{Fig1} allow us to set an upper limit on $L_{\rm c}$, of approximately 5\,pc. The simulated emissions in the plots with $L_{\rm c} \geq 10$\,pc are too asymmetric to account for the measurements.

In each panel of Fig.~\ref{Fig1}, the turbulence has a given configuration. We redo these simulations for ten different realizations of the turbulence, while keeping the same $P(k)$, and the same values of $B_{\rm rms}$ and $L_{\rm c}$. We find that the simulated emissions almost do not vary for small $L_{\rm c}$. In contrast, the shapes of the filaments at large $L_{\rm c}$ strongly depend on the realization. This is unsurprizing, because the geometry of field lines around the source is completely different from one realization to another. For small $L_{\rm c}$, magnetic field lines are too tangled on the scale of the emission for any substantial difference to be visible in the emission.

%%%%%%%%%%%%%%%%%%%%%%%%%%%%%%%%%%%%%%%%%%%%%%%%%%%%%%%%%%%%%%%%%%%%%%%%%%%%%%%%%%
\begin{figure*}
\begin{center}
\includegraphics[width=0.495\textwidth]{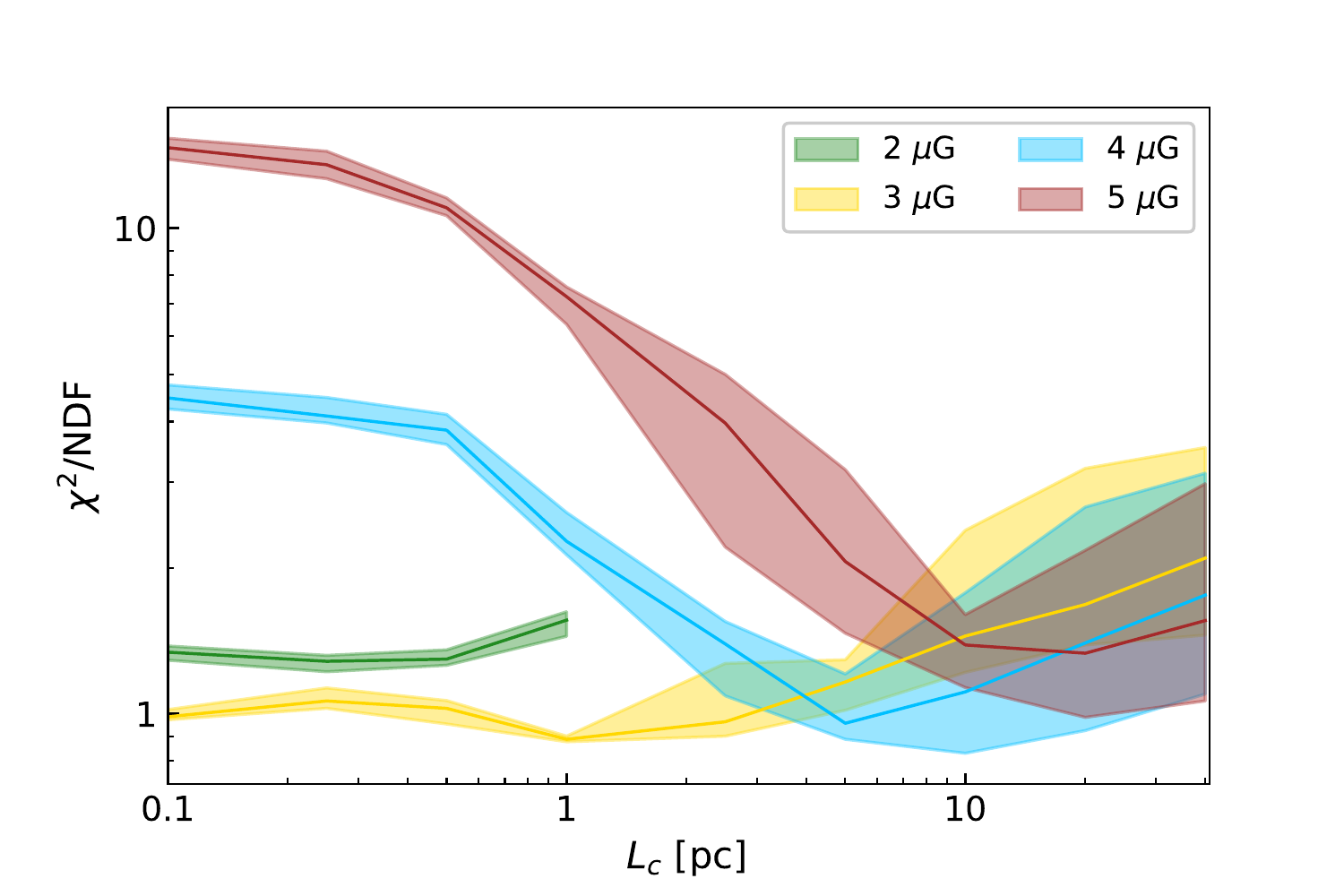}
\includegraphics[width=0.495\textwidth]{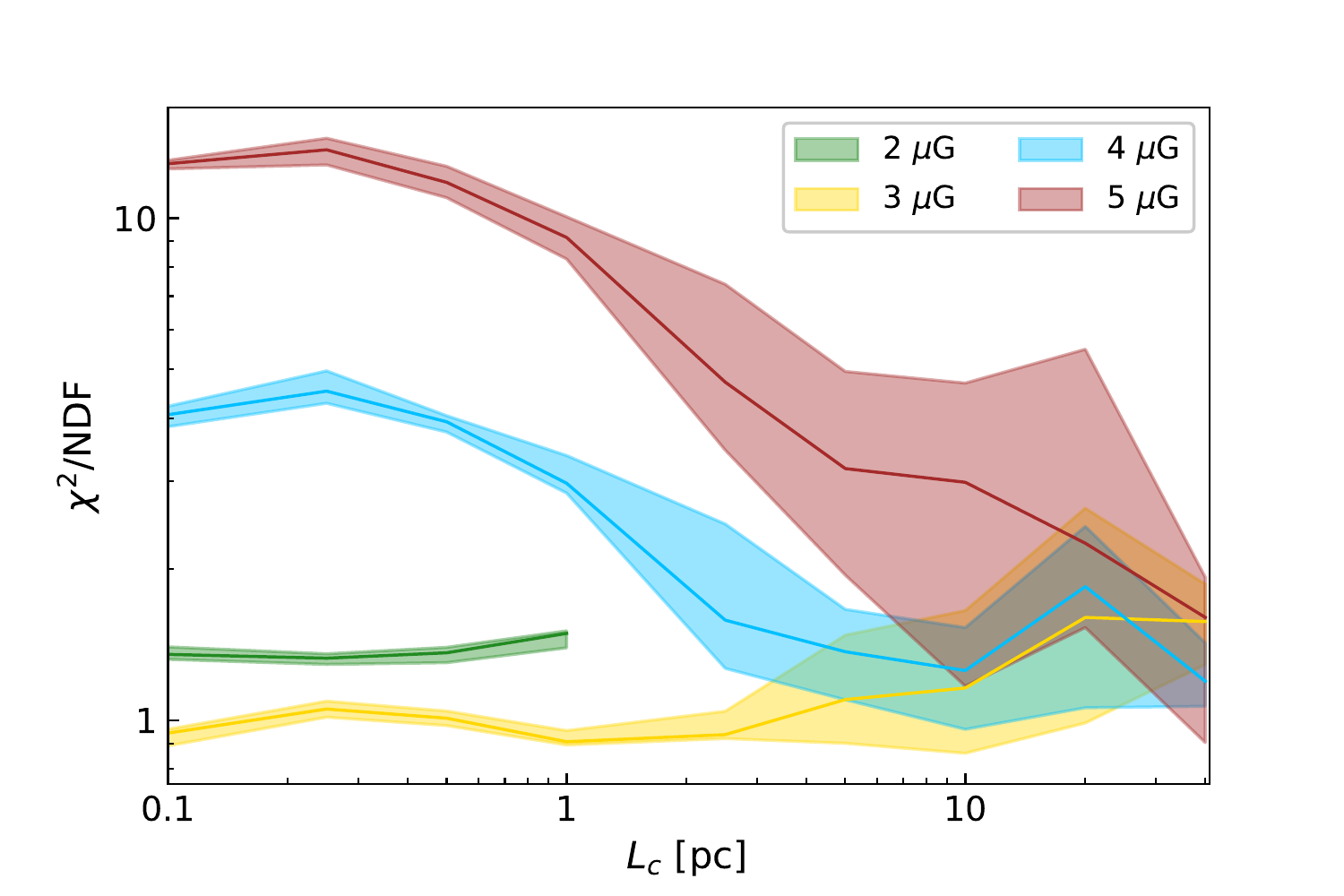}
\caption{\label{Fig2} $\chi^{2}$/ndf versus $L_{\rm c}$ for our fits to HAWC measurements from Ref.~\cite{Geminga_hawc}, for Kolmogorov (left panel) and Kraichnan (right panel) turbulence. Each colour band corresponds to a different value of $B_{\rm rms}$, see the values in the keys. The thick solid lines represent the median $\chi^{2}$/ndf, and the band width estimates the fluctuations from one realization of the turbulence to another, see the text for details.}
\end{center}
\end{figure*}
%%%%%%%%%%%%%%%%%%%%%%%%%%%%%%%%%%%%%%%%%%%%%%%%%%%%%%%%%%%%%%%%%%%%%%%%%%%%%%%%%%

For each set of parameters and each realization, we then integrate the $\gamma$-ray surface brightness over all azimuthal angles, plot it versus the angular distance to the pulsar, and fit it to HAWC measurements~\cite{Geminga_hawc}. The results for the $\chi^{2}$/ndf of these fits versus $L_{\rm c}$ are presented in Fig.~\ref{Fig2} for Kolmogorov (left panel) and Kraichnan (right) turbulence. Each line colour corresponds to a different value of $B_{\rm rms}$: 2, 3, 4, and $5\,\mu$G --- see the keys for the colour code. The width of the bands quantifies the fluctuations from one realization of the turbulence to another. The shaded region corresponds to the interval between the 18th and 82nd percentiles of all realizations. The increase of the widths of the bands with $L_{\rm c}$ is due to the reason discussed above. The thick lines inside each band correspond to the median value of $\chi^{2}$/ndf. Both for Kolmogorov and Kraichnan turbulence, the best fits are obtained for $B_{\rm rms} = 3\,\mu$G and $L_{\rm c} = 1$\,pc, where $\chi^{2}/{\rm ndf}<1$. The value of $L_{\rm c}$ at which the best fit is obtained for each value of $B_{\rm rms}$ increases with $B_{\rm rms}$. Too weak, $B_{\rm rms} \leq 2\,\mu$G, or too strong, $B_{\rm rms} \geq 5\,\mu$G, magnetic fields give bad fits. The fact that the measurements are integrated over all azimuthal angles does not allow us to take into account the constraint from the symmetry of the emission in this fit. In practice, the regions at $L_{\rm c} \geq 10$\,pc in these plots are excluded for the aforementioned reason. Finally, we note that there is no substantial difference between the two panels of Fig.~\ref{Fig2}, which shows that one cannot firmly distinguish between the two power-spectra $P(k)$ with the current measurements. However, future analyses of the dependence of the emission on $\gamma$-ray energy should be able to provide stronger constraints on the power-spectrum of the turbulence.

%%%%%%%%%%%%%%%%%%%%%%%%%%%%%%%%%%%%%%%%%%%%%%%%%%%%%%%%%%%%%%%%%%%%%%%%%%%%%%%%%%
\begin{figure}[h]
\includegraphics[width=24pc]{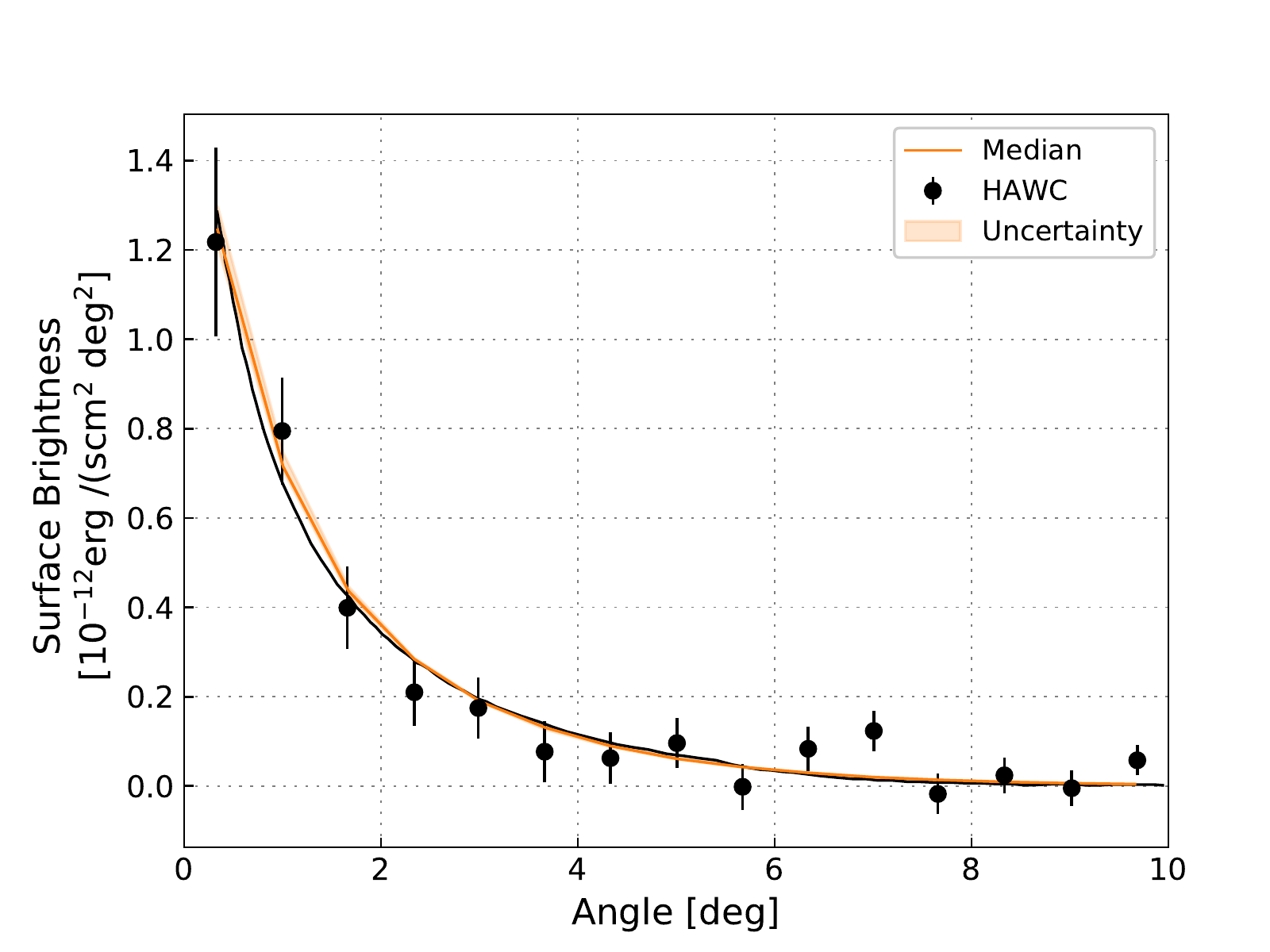}\hspace{2pc}%
\begin{minipage}[b]{14pc}\caption{\label{Fig3} $\gamma$-ray surface brightness, integrated over all azimuthal angles, versus the angular distance to the pulsar for our best fit (orange line) to HAWC measurements from Ref.~\cite{Geminga_hawc} (black dots): Kolmogorov turbulence with $B_{\rm rms} = 3\,\mu$G and $L_{\rm c}=1$\,pc. The black line corresponds to the fit from the HAWC Collaboration~\cite{Geminga_hawc}.}
\end{minipage}
\end{figure}
%%%%%%%%%%%%%%%%%%%%%%%%%%%%%%%%%%%%%%%%%%%%%%%%%%%%%%%%%%%%%%%%%%%%%%%%%%%%%%%%%%

In Fig.~\ref{Fig3}, we plot with an orange line the $\gamma$-ray surface brightness (integrated over all azimuthal angles) versus the angular distance to the pulsar for our best fit for Kolmogorov turbulence, i.e. for $B_{\rm rms} = 3\,\mu$G and $L_{\rm c} = 1$\,pc. HAWC measurements from Ref.~\cite{Geminga_hawc} is plotted with black dots, and the black line corresponds to the fit from the HAWC Collaboration~\cite{Geminga_hawc}.

%%%%%%%%%%%%%%%%%%%%%%%%%%%%%%%%%%%%%%%%%%%%%%%%%%%%%%%%%%%%%%%%%%%%%%%%%%%%%%%%%%
\section{Discussion}
%%%%%%%%%%%%%%%%%%%%%%%%%%%%%%%%%%%%%%%%%%%%%%%%%%%%%%%%%%%%%%%%%%%%%%%%%%%%%%%%%%

We have not added any regular magnetic field in the above calculations, because the apparent symmetry of the emission detected by HAWC around Geminga hints at a strongly turbulent field in this region, i.e. the amplitude of the large-scale magnetic field in this region should be smaller than that of the turbulent field. In the presence of a strong regular field, the $\gamma$-ray emission would be elongated along its direction. Such a scenario would be compatible with observations only if this field points in our direction, so that the $\gamma$-ray emission does not appear asymmetric when viewed from the Earth. Although unlikely, this is nonetheless an interesting possibility because it could help reconcile the small extent of the $\gamma$-ray emission detected by HAWC with a larger CR diffusion coefficient along the line of sight, matching the Galactic average as deduced from e.g. the boron-to-carbon ratio. If the regular field, ${\bf B}_{\rm reg}$, points towards us, the size of the $\gamma$-ray emission would appear relatively small because it would be controlled by perpendicular diffusion. In contrast, electrons would diffuse much faster along ${\bf B}_{\rm reg}$ due to parallel diffusion. The resulting asymmetry of the electron distribution around Geminga would be undetectable by an observer at Earth as long as the angle between ${\bf B}_{\rm reg}$ and the line of sight remains smaller than $\theta_{\max} \approx \tan^{-1}(\sqrt{D_{\perp}/D_{\|}})$, where $D_{\|}$ and $D_{\perp}$ are the parallel and perpendicular diffusion coefficients. Turbulence levels $B_{\rm rms}/|{\bf B_{\rm reg}}| \leq 0.5$ are needed for $D_{\|}$ to reach \lq\lq Galactic average\rq\rq\/ values with Kolmogorov turbulence, as can be seen in Fig.~3 of Ref.~\cite{Giacinti:2017dgt}. This corresponds to $D_{\|}/D_{\perp} \geq 200$, and thence $\theta_{\max} \leq 4^{\circ}$. Such a near-perfect alignment of ${\bf B}_{\rm reg}$ with the line-of-sight, and therefore the presence of such a strong regular field in this region, are quite unlikely. Also, $B_{\rm rms}/|{\bf B_{\rm reg}}|$ is thought to be greater than 1 in the Galactic disc, and the regular field is thought to follow spiral arms, which is in tension with the fact that the direction to Geminga is not aligned with that of the Orion Spur.

Finally, we note that $D_{\rm 100\,TeV}$ is not far from the Bohm value, and that the electrons may be probing turbulence generated by CRs. CR self-confinement around their sources has been studied by a number of authors, see e.g.~\cite{Ptuskin08,Malkov2013,DAngelo2016,Nava2016}, and Ref.~\cite{2018PhRvD..98f3017E} has recently suggested that it may be the reason for the low value of $D_{\rm 100\,TeV}$ measured by HAWC around Geminga. Such a scenario is possible and very interesting. Our study nonetheless shows that the current HAWC measurements can still be fitted with Kolmogorov or Kraichnan turbulence and does not require such an explanation at the present time. Radio observations suggest that the coherence length of the turbulence in the spiral arms of our Galaxy is equal to only a few parsecs, which is very close to our best fit value. See e.g. Refs.~\cite{Haverkorn:2008tb,Iacobelli2013} where the outer scale ($=5L_{\rm c}$ for Kolmogorov turbulence) is found to be $\leq 20$\,pc. Future studies of the gamma-ray emission, especially at lower energies, should be able to clarify the situation. We will address the discrepancy between $D_{\rm 100\,TeV}$ and the \lq\lq Galactic average\rq\rq\/ value from B/C measurements in a future publication.

%%%%%%%%%%%%%%%%%%%%%%%%%%%%%%%%%%%%%%%%%%%%%%%%%%%%%%%%%%%%%%%%%%%%%%%%%%%%%%%%%%
\section{Conclusions and perspectives}
%%%%%%%%%%%%%%%%%%%%%%%%%%%%%%%%%%%%%%%%%%%%%%%%%%%%%%%%%%%%%%%%%%%%%%%%%%%%%%%%%%

We have shown that the extended $\gamma$-ray emission detected by HAWC around Geminga is compatible with that from electrons propagating and cooling in Kolmogorov or Kraichnan turbulence with reasonable strengths and coherence lengths, even if the diffusion coefficient of these electrons is substantially smaller than the values usually inferred from the boron-to-carbon ratio for the Galactic average. Our best fits are obtained for turbulent fields with a root-mean-square strength of $3\,\mu$G and a coherence length of about 1\,pc. Magnetic field strengths smaller than $5\,\mu$G are favoured. Due to the lack of strong asymmetries in the observed emission, we can exclude coherence lengths greater than about 10\,pc in this $\simeq 25$\,pc-radius region around Geminga. Even though the power-spectrum of the turbulence cannot be well constrained at the present time, we expect that one could place more stringent constraints in the future by studying the energy-dependence of the morphology of the emission.

%%%%%%%%%%%%%%%%%%%%%%%%%%%%%%%%%%%%%%%%%%%%%
\ack
We thank Jim Hinton and the HAWC Collaboration for useful discussions. The research of GG was supported by a Grant from the GIF, the German-Israeli Foundation for Scientific Research and Development.
%%%%%%%%%%%%%%%%%%%%%%%%%%%%%%%%%%%%%%%%%%%%%

%%%%%%%%%%%%%%%%%%%%%%%%%%%%%%%%%%%%%%%%%%%%%
\section*{References}
\bibliography{references2}

\providecommand{\newblock}{}
\begin{thebibliography}{10}
\expandafter\ifx\csname url\endcsname\relax
  \def\url#1{{\tt #1}}\fi
\expandafter\ifx\csname urlprefix\endcsname\relax\def\urlprefix{URL }\fi
\providecommand{\eprint}[2][]{\url{#2}}

\bibitem{Geminga_hawc}
{Abeysekara} A~U {\it et al} 2017 {\em Science\/} {\bf 358} 911--914

\bibitem{2012PhRvL.108z1101G}
{Giacinti} G, {Kachelrie{\ss}} M and {Semikoz} D~V 2012 {\em Phys. Rev.
  Lett.\/} {\bf 108} 261101 (\textit{Preprint} \eprint{1204.1271})

\bibitem{2013PhRvD..88b3010G}
{Giacinti} G, {Kachelrie{\ss}} M and {Semikoz} D~V 2013 {\em Phys. Rev. D\/}
  {\bf 88} 023010 (\textit{Preprint} \eprint{1306.3209})

\bibitem{2018MNRAS.479.4526L}
{L{\'o}pez-Coto} R and {Giacinti} G 2018 {\em Mon. Not. Roy. Astron. Soc.\/}
  {\bf 479} 4526--4534

\bibitem{Diffusion_electrons_paper}
{L{\'o}pez-Coto} R, {Hahn} J, {BenZvi} S, {Dingus} B, {Hinton} J, {Nisa} M~U,
  {Parsons} R~D, {Greus} F~S, {Zhang} H and {Zhou} H 2018 {\em Astropart.
  Phys.\/} {\bf 102} 1--11 (\textit{Preprint} \eprint{1709.07653})

\bibitem{Moderski05}
{Moderski} R, {Sikora} M, {Coppi} P~S and {Aharonian} F 2005 {\em Mon. Not.
  Roy. Astron. Soc.\/} {\bf 363} 954--966 (\textit{Preprint}
  \eprint{astro-ph/0504388})

\bibitem{Giacinti2012}
{Giacinti} G, {Kachelrie{\ss}} M, {Semikoz} D~V and {Sigl} G 2012 {\em Journ.
  Cosmol. Astropart. Phys.\/} {\bf 7} 031 (\textit{Preprint}
  \eprint{1112.5599})

\bibitem{GAMERA}
Hahn J 2015 {\em Proceedings of the 34th International Cosmic Ray Conference,
  id 917\/}

\bibitem{Giacinti:2017dgt}
{Giacinti} G, {Kachelriess} M and {Semikoz} D~V 2018 {\em J. Cosmol. Astropart.
  Phys.\/} {\bf 7} 051 (\textit{Preprint} \eprint{1710.08205})

\bibitem{Ptuskin08}
{Ptuskin} V~S, {Zirakashvili} V~N and {Plesser} A~A 2008 {\em Advances in Space
  Research\/} {\bf 42} 486--490

\bibitem{Malkov2013}
{Malkov} M~A, {Diamond} P~H, {Sagdeev} R~Z, {Aharonian} F~A and {Moskalenko}
  I~V 2013 {\em Astrophys. J.\/} {\bf 768} 73 (\textit{Preprint}
  \eprint{1207.4728})

\bibitem{DAngelo2016}
{D'Angelo} M, {Blasi} P and {Amato} E 2016 {\em Phys. Rev. D\/} {\bf 94} 083003
  (\textit{Preprint} \eprint{1512.05000})

\bibitem{Nava2016}
{Nava} L, {Gabici} S, {Marcowith} A, {Morlino} G and {Ptuskin} V~S 2016 {\em
  Mon. Not. Roy. Astron. Soc.\/} {\bf 461} 3552--3562 (\textit{Preprint}
  \eprint{1606.06902})

\bibitem{2018PhRvD..98f3017E}
{Evoli} C, {Linden} T and {Morlino} G 2018 {\em Phys. Rev. D\/} {\bf 98} 063017
  (\textit{Preprint} \eprint{1807.09263})

\bibitem{Haverkorn:2008tb}
{Haverkorn} M, {Brown} J~C, {Gaensler} B~M and {McClure-Griffiths} N~M 2008
  {\em Astrophys. J.\/} {\bf 680} 362-370 (\textit{Preprint}
  \eprint{0802.2740})

\bibitem{Iacobelli2013}
{Iacobelli} M {\it et al} 2013 {\em Astron. Astrophys.\/} {\bf 558} A72
  (\textit{Preprint} \eprint{1308.2804})

\end{thebibliography}
%%%%%%%%%%%%%%%%%%%%%%%%%%%%%%%%%%%%%%%%%%%%%

\end{document}